\documentclass[12pt]{article}
\usepackage{subeqn}
\usepackage{epsfig}
\newcommand{\be}{\begin{equation}}
\newcommand{\ee}{\end{equation}}

\newcommand{\no}{\noindent}
\newcommand{\ce}{\begin{center}}
\newcommand{\nc}{\end{center}}

\makeatletter
\@addtoreset{equation}{section}
\makeatother

\def\sqr#1#2{{\vcenter{\vbox{\hrule height.#2pt
 \hbox{\vrule width.#2pt height#1pt \kern#1pt
 \vrule width.#2pt} \hrule height.#2pt}}}}

\def\operp{\hbox{${\kern+.25em{\bigcirc}
\kern-.85em\bot\kern+.85em\kern-.25em}$}}

\def\lsim{\;\raise0.3ex\hbox{$<$\kern-0.75em\raise-1.1ex\hbox{$\sim$}}\;}
\def\gsim{\;\raise0.3ex\hbox{$>$\kern-0.75em\raise-1.1ex\hbox{$\sim$}}\;}
\def\no{\noindent}

\def\ce{\centerline}
\def\ve{\vfill\eject}
\def\rdots{\mathinner{\mkern1mu\raise1pt\vbox{\kern7pt\hbox{.}}\mkern2mu
 \raise4pt\hbox{.}\mkern2mu\raise7pt\hbox{.}\mkern1mu}}

\def\e e{$e^+ e^-$ }

\begin{document}

\ce{\bf COLORED PREONS}

\vskip.3cm

\ce{\it Robert J. Finkelstein}
\vskip.3cm

\ce{Department of Physics and Astronomy}
\ce{University of California, Los Angeles, CA 90095-1547}

\vskip1.0cm

\no{\bf Abstract.}  Previous studies have suggested complementary 
models of the elementary particles as (a) quantum knots and (b) preonic
nuclei that are field and particle descriptions, respectively, of the
same particles.  This earlier work, carried out in the context of
standard electroweak $(SU(2)\times U(1))$ physics, is here extended
to the strong interactions by the introduction of color $(SU(3))$
charges.

\vskip5.0cm

\no UCLA/08/TEP/34

\ve

\section{Introduction}

The study of elementary particles as
quantum knots in the context of electroweak physics leads
in a natural way to a dual description of the elementary
fermions as 
composite structures composed of three preons.$^1$  We are here
interested in investigating the compatibility of this structure with
the strong interactions.

It had been previously shown$^{2,3,4}$ 
that the four classes of elementary
fermions \break $(e,\mu,\tau;\nu_e,\nu_\mu,\nu_\tau;d,s,b;u,c,t)$, 
carrying the
electroweak quantum numbers $(t,t_3,t_0)$, may be regarded as
quantum knots described by irreducible representations, 
${\cal{D}}^{3t}_{-3t_3-3t_0}(q|abcd)$, of $SL_q(2)$ in the manner shown
in Table 1.

\vskip.3cm

\ce{\bf Table 1.}

\[
\begin{array}{cccccc}
\underline{f_1f_2f_3} & \underline{t} & \underline{t_3} &
\underline{t_0} & \underline{Q} & \underline{{\cal{D}}^{3t}_{-3t_3-3t_0}
} \\
e\mu\tau & \frac{1}{2} & -\frac{1}{2} & -\frac{1}{2} & -e &
{\cal{D}}^{3/2}_{\frac{3}{2}\frac{3}{2}} \\
\nu_e\nu_\mu\nu_\tau & \frac{1}{2} & \frac{1}{2} & -\frac{1}{2} & 0 &
{\cal{D}}^{3/2}_{-\frac{3}{2}\frac{3}{2}} \\
dsb & \frac{1}{2} & -\frac{1}{2} & \frac{1}{6} & -\frac{1}{3}e &
{\cal{D}}^{3/2}_{\frac{3}{2}-\frac{1}{2}} \\
uct & \frac{1}{2} & \frac{1}{2} & \frac{1}{6} & \frac{2}{3}e &
{\cal{D}}^{3/2}_{-\frac{3}{2}-\frac{1}{2}} \\
\end{array}
\]

\no Here $Q = e(t_3+t_0)$ is the expression for the electric charge
in the standard theory.

The ${\cal{D}}^j_{mm^\prime}(q|abcd)$ are functions lying in the
$SU_q(2)$ algebra and are given explicitly by$^1$
\be
{\cal{D}}^j_{mm^\prime}(q|abcd) = \sum_{\scriptstyle s\leq n_+
\atop\scriptstyle t\leq n_-} {\cal{A}}^j_{mm^\prime}(q,s,t)
\delta(s+t,n_+^\prime)a^sb^{n_+-s}c^td^{n_--t}
\ee
where 
\begin{eqnarray*}
n_\pm &=& j\pm m \\ n_\pm^\prime &=& j\pm m^\prime
\end{eqnarray*}
and the arguments of ${\cal{D}}^j_{mm^\prime}$ obey the algebra of
$SL_q(2)$, the ``knot algebra", as follows:
\[
\left.
\begin{array}{l}
ab = qba \\ ac = qca
\end{array} \right. \quad \left.
\begin{array}{l}
bd = qdb \\ cd = qdc
\end{array} \right. \quad \left.
\begin{array}{l}
bc = cb \\ \hfil 
\end{array} \right. \quad \left.
\begin{array}{l}
ad-qbc=1 \\ da-q_1cb=1 
\end{array} \right.  \qquad
q_1=q^{-1}  \hskip1.0cm (A)
\]

The ${\cal{A}}^j_{mm^\prime}(q,s,t)$ are numerical functions that do not concern us here.

By (1.1)
the explicit form of the different ${\cal{D}}^{3t}_{-3t_3-3t_0}
(q|abcd)$ shown in Table 1 are given in Table 2.  Ignoring the
numerical coefficients we see that these functions are simple
monomials carrying the quantum numbers that are correct for the
different fermion families.

\vskip.3cm

\ce{\bf Table 2.}
\[
\begin{array}{cc}
& \underline{{\cal{D}}^{3t}_{-3t_3-3t_0}(q|a,b,c,d)} \\
e\mu\tau & {\cal{D}}^{3/2}_{\frac{3}{2}\frac{3}{2}} = a^3 \\
\nu_e\nu_\mu\nu_\tau & {\cal{D}}^{3/2}_{-\frac{3}{2}\frac{3}{2}} = c^3 \\
dsb & {\cal{D}}^{3/2}_{\frac{3}{2}-\frac{1}{2}}=ab^2 \\
uct & {\cal{D}}^{3/2}_{-\frac{3}{2}-\frac{1}{2}}=cd^2
\end{array}
\]
It has been previously noted that the state functions of these four
quantum knots may be interpreted as composite field operators where
the $(a,b,c,d)$ are preon creation operators, carrying the values of
$(t,t_3,t_0)$ as shown in Table 3.$^1$

\vskip.3cm

\ce{\bf Table 3.}
\[
\begin{array}{cccccc}
\underline{\rm Preon} & \underline{t} & \underline{t_3} & \underline{t_0} & \underline{Q} & \underline{{\cal{D}}^{3t}_{-3t_3-3t_0}} \\
a & \frac{1}{6} & -\frac{1}{6} & -\frac{1}{6} & -\frac{e}{3} &
{\cal{D}}^{1/2}_{\frac{1}{2}\frac{1}{2}} \\
b & \frac{1}{6} & -\frac{1}{6} & \frac{1}{6} & 0 &
{\cal{D}}^{1/2}_{\frac{1}{2}-\frac{1}{2}} \\
c & \frac{1}{6} & \frac{1}{6} & -\frac{1}{6} & 0 &
{\cal{D}}^{1/2}_{-\frac{1}{2}\frac{1}{2}} \\
d & \frac{1}{6} & \frac{1}{6} & \frac{1}{6} & \frac{e}{3} &
{\cal{D}}^{1/2}_{-\frac{1}{2}-\frac{1}{2}} 
\end{array}
\]
In Table 3 the relation $Q=e(t_3+t_0)$ of the standard theory
is retained.

By Table 3 we see that
$a$ and $d$ behave as creation operators for the charged $(-e/3)$
preon and $(e/3)$ antipreon respectively while $b$ and $c$ are neutral
with opposite values of both $t_3$ and $t_0$.

The four preon operators are elements of the fundamental representation
of $SL_q(2)$ as follows:
\[
{\cal{D}}^{1/2}_{mm^\prime}: \qquad  
\begin{array}{c|cc}
{}_m\backslash{}^{m'} & \frac{1}{2} & -\frac{1}{2} \\ \hline 
\frac{1}{2} & a & b \\
-\frac{1}{2} & c & d
\end{array}
\]

The four elementary fermion operators are elements of the
${\cal{D}}^{3/2}_{mm^\prime}$ representation.  In both the $j=1/2$ 
case and the $j=3/2$ case we set
\begin{eqnarray*}
j &=& 3t \\ m &=& -3t_3 \\
m^\prime &=& -3t_0
\end{eqnarray*}

In this way $t,t_3$, and $t_0$ are all defined with respect to the
$SL_q(2)$ algebra while in the standard theory $t,t_3$, and $t_0$ are
defined by the $SU(2)\times U(1)$ algebra.  In the case that
$j=3/2$ the two definitions agree.

According to these same ideas there should also be bosonic preons
corresponding to the adjoint representation $D^1_{mm^\prime}$.  The
operators for these particles are products of two fermionic preon
operators just as the operators for leptons, neutrinos, and quarks are
products of three fermionic preon operators.

\vskip.5cm

\section{Gluon Charge}

The previous considerations are based on electroweak physics.
To describe the strong interactions it is necessary according to
standard theory to introduce $SU(3)$ charge.  We shall 
therefore assume that
each of the four preon operators appears in triplicate
$(a_i,b_i,c_i,d_i)$ where $i=R,Y,G$, without changing the algebra
$(A)$.  These colored preon operators provide a basis for the
fundamental representation of $SU(3)$ just as the colored quark
operators do in standard theory.  To adapt the electroweak operators
to the requirements of gluon fields we make the following
replacements:
\begin{eqnarray}
{\rm leptons}:  & &a^3 \to \epsilon^{ijk}a_ia_ja_k \\
{\rm neutrinos}:  & & c^3 \to \epsilon^{ijk}c_ic_jc_k \\
{\rm down~quarks}:  & & ab^2 \to a_ig^{jk}b_jb_k \equiv
a_i(b^kb_k) \\
{\rm up~quarks}:  & & cd^2 \to c_ig^{jk}d_jd_k \equiv
c_i(d^kd_k)
\end{eqnarray}
where $g^{jk}$ is the group metric of $SU(3)$ and $(i,j,k) =
(R,Y,G)$ and $(a_ib_ic_id_i)$ are creation operators for colored
preons.  
Then the leptons and neutrinos are color singlets while
the quark states correspond to the fundamental representation of
$SU(3)$, as required by standard theory.

\vskip.5cm

\section{The Complementary Models}

We have ascribed to the quantum knot the state function
${\cal{D}}^j_{mm^\prime}(q|abcd)$, an irreducible representation of the
knot algebra $SL_q(2)$, where the indices $j=\frac{N}{2}$, $m=\frac{w}{2}$, $m^\prime = \frac{r+1}{2}$ are restricted to values of $(N,w,r)$
allowed by the classical (geometrical) knots.  The quantum knots have
more degrees of freedom than their classical images with the
consequence that two quantum knots may be distinguishable when their
classical images are not.  In particular there are four
distinguishable quantum trefoils with $(w,r) = (\pm 3,\pm 2)$ but only
two of their classical images $(w,r) = (\pm 3,2)$ are 
topologically different.  In
the physical application $(w,r) = (\pm 3,2)$ describe the leptons and
neutrinos while $(w,r) = (\pm 3,-2)$ describe the two varieties of
quarks, i.e., the two additional quantum knots are required to permit
the description of colored fermions.

These considerations have led us to two complementary models of the
elementary particles, namely
\begin{description}
\item{(a)} quantum knots
\item{(b)} preon structures
\end{description}
that are the field and particle descriptions of the same particles.
The correspondence may be expressed by the following relations$^1$
\begin{eqnarray}
w &=& n_a-n_d+n_b-n_c(=2m=-6t_3) \\
r+1 &=& n_a-n_d-n_b+n_c(=2m^\prime=-6t_0) \\
N &=& n_a+n_b+n_c+n_d(=2j=6t)
\end{eqnarray}
Here $(N,w,r)$ describe the number of crossings, the writhe and the
rotation of the particle regarded as a quantum knot of field while
$(n_a,n_b,n_c,n_d)$ record the number of $(a,b,c,d)$ preons in the
dual description of the same structure.  We have also described this
particle by
\be
{\cal{D}}^j_{mm^\prime} = {\cal{D}}^{3t}_{-3t_3-3t_0} =
{\cal{D}}^{N/2}_{\frac{w}{2}\frac{r+1}{2}}
\ee
as indicated in (3.1)-(3.3).

The knot $(N,w,r)$ and the preon $(n_a,n_b,n_c,n_d)$ descriptions share
the same representation of $SU_q(2)$ as follows.

In terms of $(N,w,r)$ one has $D^j_{mm^\prime} = 
D^{N/2}_{\frac{w}{2}\frac{r+1}{2}}$ where
\be
D^{N/2}_{\frac{w}{2}\frac{r+1}{2}}(q|abcd) = \left[
\frac{\langle n_+^\prime\rangle!\langle n_-^\prime\rangle!}
{\langle n_+\rangle!\langle n_-\rangle!}\right]^{1/2}
\sum_{\scriptstyle 0\leq s\leq n_+\atop\scriptstyle 0\leq t\leq n_-}
\left\langle\matrix{n_+\cr s}\right\rangle_{q_1}
\left\langle\matrix{n_-\cr t}\right\rangle_{q_1}
\delta(s+t,n_+^\prime)a^sb^{n_+-s}c^td^{n_--t}
\ee
and again in terms of $(N,w,r)$
\begin{eqnarray}
n_\pm &=& \frac{1}{2}[N\pm w] \\
n_\pm^\prime &=& \frac{1}{2}[N\pm(r+1)]
\end{eqnarray}
The complementary description expressed
in terms of the population numbers $(n_a,n_b,n_c,n_d)$ is
\be
D^j_{mm^\prime} = {\cal{D}}^N_{\nu_a\nu_b}
\ee
where
\be
{\cal{D}}^N_{\nu_a\nu_b} = \left[\frac{\langle n_a+n_c\rangle!
\langle n_b+n_d\rangle!}{\langle n_a+n_b\rangle!
\langle n_a+n_d\rangle!}\right]^{1/2} 
\sum_{\scriptstyle N\geq n_a,n_b\geq 0\atop\scriptstyle
N\geq n_c,n_d\geq 0}
\left\langle\matrix{n_a+n_b\cr n_a}\right\rangle_{q_1}
\left\langle\matrix{n_c+n_d\cr n_c}\right\rangle_{q_1}
a^{n_a}b^{n_b}c^{n_c}d^{n_d}
\ee
The limits on $\sum$, literally translated from $D^j_{mm^\prime}$ are shown in the expression for ${\cal{D}}^N_{\nu_a\nu_b}$ but these
limits simply describe the requirement that all population numbers,
$n_i$ satisfy $N \geq n_i \geq 0$.

Here the exponents $(n_a,n_b,n_c,n_d)$ of $(abcd)$ are
\be
\begin{array}{ll}
n_a = s & n_b = n_+-s \\
n_c = t & n_d = n_--t
\end{array}
\ee
They are related to $(n_+,n_-,n^\prime_+,n_-^\prime)$ by
\be
\begin{array}{ll}
n_+ = n_a+n_b & n_+^\prime = n_a+n_c \\
n_- = n_c+n_d & n_-^\prime = n_b+n_d
\end{array}
\ee
Since $a$ and $d$ have opposite charge and hypercharge, while $b$ and
$c$ are neutral with opposite hypercharge, we may define the ``preon
numbers" $\nu_a$ and $\nu_b$ as follows
\be
\begin{array}{rcl}
\nu_a &=& n_a-n_d \\
\nu_b &=& n_b-n_c
\end{array}
\ee
Then by (3.1) and (3.2)
\be
\begin{array}{rcl}
\nu_a + \nu_b &=& 2m=w \\
\nu_a-\nu_b &=& 2m^\prime = r+1
\end{array}
\ee
Then the conservation of the writhe and rotation implies the conservation of the preon numbers $\nu_a$ and $\nu_b$.
According to Table 2 the trefoil solutions of (3.1)-(3.3) are given
in Table 4.

\vskip.3cm

\ce{\bf Table 4.}

\[
\begin{array}{ccccc}
\quad &  \underline{n_a} \quad & \underline{n_b} \quad & \underline{n_c} \quad & \underline{n_d} \\
\ell \quad & 3 \quad & 0  \quad & 0 \quad & 0 \\
\nu \quad & 0 \quad & 0 \quad & 3 \quad & 0 \\
d \quad & 1 \quad & 2 \quad & 0 \quad & 0 \\
u \quad & 0 \quad & 0 \quad & 1 \quad & 2
\end{array}
\]

Since the number of crossings equals the number of preons, one may
speculate that there is one preon at each crossing if both preons and
crossings are considered pointlike. 
If the pointlike crossings are
labelled $(\vec x_1\vec x_2\vec x_3)$, then by (2.1)-(2.4) the wave
functions of the trefoils representing leptons $(\ell)$, neutrinos
$(\nu)$, down quarks $(d)$, up quarks $(u)$ are as follows:
\begin{eqnarray}
\Psi_\ell(\vec x_1\vec x_2\vec x_3) &=& \epsilon^{ijk}\psi_i
(a|\vec x_1)\psi_j(a|\vec x_2)\psi_k(a|\vec x_3) \\ 
\Psi_\nu(\vec x_1\vec x_2\vec x_3) &=& \epsilon^{ijk}
\psi_i(c|\vec x_1)\psi_j(c|\vec x_2)\psi_k(c|\vec x_3) \\
\Psi_d(\vec x_1\vec x_2\vec x_3) &=& \psi_i(a|\vec x_1)
\psi^j(b|\vec x_2)\psi_j(b|\vec x_3) \\
\Psi_u(\vec x_1\vec x_2\vec x_3) &=& \psi_i(c|\vec x_1)
\psi^j(d|\vec x_2)\psi_j(d|\vec x_3)
\end{eqnarray}
where $i=(R,Y,G)$ and $\psi_i(a|\vec x) \ldots \psi_i(d|\vec x)$
are colored $\delta$-like functions localizing the preons at the
crossings.

Then the wave function of a lepton describes a singlet trefoil
particle containing three preons of charge $(-e/3)$ and
hypercharge $(-e/6)$.  The corresponding characterization of a
neutrino describes a singlet trefoil containing three neutral
preons of hypercharge $(-e/6)$.

The wave function of a down quark describes a colored trefoil particle
containing one $a$-preon with charge $(-e/3)$ and hypercharge
$(-e/6)$ and two neutral $b$-preons with hypercharge $(e/6)$.  The
corresponding characterization of an up-quark describes a colored
trefoil containing two charged $d$-preons with charges $(e/3)$
and hypercharge $(e/6)$, and one neutral $c$-preon with hypercharge
$(-e/6)$.

This hypothetical structure would
be held together by the fields connecting the charged preons.  Since
the preons carry both electric charge and hypercharge as well as color
charge and color hypercharge, the total Lagrangian that determines
these fields, and therefore the preon dynamics, would then be 
determined by the sum
of two non-Abelian Lagrangians, one describing standard electroweak
based on local $SU(2)\times U(1)$ and the other describing standard
chromodynamics based on local $SU(3)$. Here 
we may assume that the $SU(2)\times U(1)$ and
the $SU(3)$ gauge bosons are also quantum knots described by the
adjoint representation ${\cal{D}}^1_{mm^\prime}(q|abcd)$ of global
$SU_q(2)$.  In this way the preons might play a similar role with
respect to the tripreon (the basic fermion) that the quarks play
with respect to the triquark (the hadron).  A search for this kind of
substructure depends critically on the mass of the conjectured preon
and about which nothing can unfortunately 
be said with any confidence.
Nevertheless we shall try to extend the Higgs idea to a consideration
of the mass of the preon.

\vskip.5cm

\section{Mass of Preons}

Since the preons are necessarily assumed to be pointlike, they must be
very heavy.  Let us assume that the mass of the preon is computed in
the same way as we have computed
the mass of the elementary fermions, i.e., by 
adopting the mass terms of the standard theory, namely$^1$
\be
{\cal{M}} = \bar L\varphi R + \bar R\varphi L
\ee
where $L$ and $R$ are left and right chiral spinors and $\varphi$ is
the Higgs scalar.

We shall assign a $SU_q(2)$ singlet structure to $\varphi$ and the 
preon representation $D^{1/2}_{mm^\prime}$ to both $L$ and $R$.  Then
we substitute for $L$ and $R$ as follows:
\begin{eqnarray}
L \to \chi_L D^{1/2}_{mm^\prime}|0\rangle \\
R \to \chi_R D^{1/2}_{mm^\prime}|0\rangle
\end{eqnarray}
where $\chi_L$ and $\chi_R$ are the standard fermionic fields and
$D^{1/2}_{mm^\prime}|0\rangle$ describes the internal structure of 
the preons.  Here $|0\rangle$ is the ground state of the $SU_q(2)$
algebra.  Then
\begin{eqnarray}
{\cal{M}} &\to& \langle 0|\bar D^{1/2}_{mm^\prime}D^{1/2}_{mm^\prime}
|0\rangle(\bar\chi_L\varphi\chi_R + \bar\chi_R\varphi\chi_L) \\
& &\mbox{}=M(m,m^\prime)\bar\chi\chi
\end{eqnarray}
where the mass is
\be
M(m,m^\prime) = \rho(m,m^\prime)\langle 0|\bar D^{1/2}_{mm^\prime}
D^{1/2}_{mm^\prime}|0\rangle
\ee
where $\rho(m,m^\prime)$ is a local minimum of the Higgs potential.  We
shall assume that there are 4 local minima of the Higgs potential, 
namely
\be
(m,m^\prime) = \left(\pm\frac{1}{2},\pm\frac{1}{2}\right)
\ee
For example, the mass of the $D^{1/2}_{\frac{1}{2}\frac{1}{2}}$ preon
is
\be
M\left(\frac{1}{2},\frac{1}{2}\right) = \rho\left(\frac{1}{2},
\frac{1}{2}\right) \langle 0|\bar aa|0\rangle
\ee
The mass of the electron computed in the same way is
\be
M\left(\frac{3}{2},\frac{3}{2}\right) = 
\rho\left(\frac{3}{2},\frac{3}{2}\right) \langle 0|\bar a^3a^3|0\rangle
\ee
Then the ratio of the preon mass $(m_p)$ to the electron mass
$(m_e)$ is
\be
\frac{m_p}{m_e} = \frac{\rho\left(\frac{1}{2},\frac{1}{2}\right)}
{\rho\left(\frac{3}{2},\frac{3}{2}\right)}
\frac{\langle 0|\bar aa|0\rangle}{\langle 0|\bar a^3a^3|0\rangle}
\ee
The factor $\frac{\langle 0|\bar aa|0\rangle}{\langle 0|\bar a^3a^3|
0\rangle}$ is a simple function of $q$ and $|\beta|$.

If the ratio of the Compton wavelengths is greater than unity, i.e.,
if
\be
\frac{\lambda_e}{\lambda_p} = \frac{m_p}{m_e} = 
\frac{\rho\left(\frac{1}{2},\frac{1}{2}\right)}
{\rho\left(\frac{3}{2},\frac{3}{2}\right)}
\frac{\langle 0|\bar aa|0\rangle}{\langle 0|\bar a^3a^3|0\rangle}
 > 1
\ee
then the mass of the preon is compatible with the pointlike extension
of the electron.  Under the foregoing assumptions this preon model would
require several minima in the Higgs potential satisfying relations
such as the following:
\be
\rho\left(\frac{1}{2},\frac{1}{2}\right) >
\frac{\langle 0|\bar a^3a^3|0\rangle}{\langle 0|\bar aa|0\rangle}
\rho\left(\frac{3}{2},\frac{3}{2}\right)
\ee
where $\rho\left(\frac{3}{2},\frac{3}{2}\right)$ and
$\rho\left(\frac{1}{2},\frac{1}{2}\right)$ are local minima of the Higgs
potential.  Here $\rho\left(\frac{3}{2},\frac{3}{2}\right)$ is in the
estimated neighborhood of the Higgs mass that appears in the
standard theory.

There is a stronger limit on $m_p$ if $\lambda_p$ is of the order of
$10^{-16}$ cm as suggested by scattering experiments.  In any case
$m_p$ is much greater than the masses of the leptons, neutrinos and
quarks.  According to any nuclear physics model the binding energy
must therefore almost totally compensate the sum of the masses of the
three preons.  On the other hand, according to the model underlying
(4.12), the neutrino, lepton, quark, and preon masses are 
generated by the symmetry breaking Higgs potential.

\vskip.5cm

\no {\bf References}

\begin{enumerate}
\item R. J. Finkelstein, hep-th/08063105.
\item R. J. Finkelstein, Int. J. Mod. Phys. A{\bf 20}, 6481 (2005).
\item A. C.. Cadavid and R. J. Finkelstein, {\it ibid.} A{\bf 25},
4264 (2006).
\item R. J. Finkelstein, {\it ibid.} A{\bf 22}, 4467 (2007).
\end{enumerate}

\end{document}